\documentclass{biophys-new}
\usepackage[utf8]{inputenc}
\usepackage{graphicx}
\usepackage[colorlinks,allcolors=cyan!70!black]{hyperref}
\usepackage{lipsum}

\usepackage{graphicx} 

\usepackage{amsmath}
\usepackage{siunitx}

\usepackage{makecell}
\usepackage{colortbl}

\title{Three-Body Interactions of Lipid Membrane-Deforming Colloidal Spheres}
\runningtitle{Biophysical Journal Template} 

\author[1]{Ali Azadbakht}
\author[2,3]{Billie Meadowcroft}
\author[2]{Juraj M\'ajek}
\author[2]{An\dj ela \v{S}ari\'c}
\author[1*]{Daniela J. Kraft}
\runningauthor{Azadbakht et al.} 
\affil[1]{Soft Matter Physics, Huygens-Kamerlingh Onnes Laboratory, Leiden University, PO Box 9504, 2300 RA Leiden, the Netherlands}
\affil[2]{Institute of Science and Technology Austria, 3400 Klosterneuburg, Austria}
\affil[3]{Department of Physics and Astronomy, Institute for the Physics of Living Systems, University College London, London WC1E 6BT, United Kingdom}

\corrauthor[*]{kraft@physics.leidenuniv.nl}
\papertype{Article}

\begin{document}

\begin{frontmatter}
\begin{abstract}
Many cell functions require a concerted effort from multiple membrane proteins, for example, for signaling, cell division, and endocytosis. One contribution to their successful self-organization stems from the membrane deformations that these proteins induce. While the pairwise interaction potential of two membrane deforming spheres has recently been measured, membrane-deformation induced interactions have been predicted to be non-additive and hence their collective behavior cannot be deduced from this measurement. We here employ a colloidal model system consisting of adhesive spheres and giant unilamellar vesicles to test these predictions by measuring the interaction potential of the simplest case of three membrane-deforming spherical particles. We quantify their interactions and arrangements and for the first time experimentally confirm and quantify the non-additive nature of membrane-deformation induced interactions. We furthermore conclude that there exist two favorable configurations on the membrane: (1) a linear, and (2) a triangular arrangement of the three spheres. Using Monte Carlo simulation we corroborate the experimentally observed energy minima and identify a lowering of the membrane deformation as the cause for the observed configurations. The high symmetry of the preferred arrangements for three particles suggests that arrangements of many membrane-deforming objects might follow simple rules.  
\end{abstract}

\begin{sigstatement}
Lipid membrane deforming objects, such as proteins, can interact through the membrane curvature they impose.  These interactions have been suggested to be non-additive, that is, one cannot extrapolate from the interaction between two objects the interactions between three or more such objects. In addition, the governing equations are so involved that there are only few and contradicting theoretical and numerical predictions. In this manuscript, this interaction is quantified for the first time for three spherically symmetric deformations on spherical membranes through a series of experiments and Monte Carlo simulations. We find two preferred states: a linear arrangement for smaller distances and an equilateral triangle for slightly larger interparticle distances.   
\end{sigstatement}
\end{frontmatter}

\section*{Introduction}
Eukaryotic cells are surrounded by a phospholipid plasma membrane with proteins that make up about half of their membrane\cite{AlbertsB2003}. The organization of these proteins into larger complexes not only plays a vital role in many functional membrane processes such as membrane trafficking, cell division, and endocytosis, but may also cause disease~\cite{Irvine2008,Ashraf2014,PFEFFERKORN2012162}. A precise understanding of the interactions between many proteins  is necessary to unravel how the assembly pathway leads to functional or malignant complexes.

However, protein interactions are by no means simple. The proteins feature a complex shape and posses different physicochemical properties which may lead to Van der Waals interactions~\cite{Roth1996}, electrostatics~\cite{Zhang2011}, and hydrophobic and hydrophilic interactions~\cite{Killian1998}. In addition, they may be interacting with or embedded in the cell membrane, an environment that is composed of various lipids, integral, peripheral, and cytoskeleton proteins, and which can induce interactions due to Casimir-like thermal undulations of the membrane~\cite{Podgornika1996,Pezeshkian2017}, and membrane deformations~\cite{Goulian1993}. This complexity makes it difficult to selectively measure the contribution from a single interaction, such as the one stemming from membrane-deformations. In addition, the small size of proteins makes it challenging to gain dynamic information about their arrangements and hence interactions. 

For some interactions, such as electrostatics, a well-understood and widely tested theoretical framework is available that can be used to confidently make predictions. For membrane-deformation based interactions, however, the governing equations are so complex that they preclude full analytical solutions. To simplify the problem, the lipid membrane is often modelled as a tensionless flat sheet that is being deformed by  objects with simple shapes, such as cones, disks, or spheres. Using such an approach, it was predicted that membrane-bending mediated interactions can provide a significant and possibly even dominant contribution to the overall interaction~\cite{Goulian1993, Park1996, Kim1998}. However, theoretical analyses are restricted to linear approximations of the involved equations~\cite{Goulian1993, yolcu2014a}, making them only valid for the small perturbations regime where the membrane shape is not significantly deformed.

More recently, colloidal particles that deform lipid membranes have been employed to serve as models for quantitatively investigating membrane-mediated interactions in experiments. Using such a system, two-body interactions stemming from membrane deformations were measured for the first time and found to be attractive. However, understanding the arrangement of many particles in such systems is challenging due to the non-pairwise additive nature of these interactions~\cite{Kim1998,Dommersnes1999}. Analytical attempts to apply a first-order approximation of the indentation for three membrane-deforming objects predicted that the equilateral triangle is the most favorable arrangement, while a linear arrangement has been suggested to occur occasionally~\cite{Park1996,Yolcu2012}. However, these approaches also predicted a repulsion between symmetric membrane-deforming objects,\cite{Goulian1993,Dommersnes1998} which is in contrast to recent experiments on model systems and simulations that found an attraction for membrane-wrapped colloidal particles, which strongly deform the membrane~\cite{Reynwar2007a, VanDerWel2016, Sarfati2016}.

Early computer simulations on flat sheets showed that a linear superposition of the curvatures induced by multiple membrane deforming objects does not correctly reflect the energy between them~\cite{Kim1998}. For spherical membrane-deforming particles whose degree of wrapping can be varied, it was found that the membrane bending energy and binding energy of the particles to the membrane determine their assembled state, which can be a linear, hexagonally ordered, or arrested aggregate~\cite{Saric2012}. Interestingly, due to the many body-interactions, the linear arrangement was found to be favored for a wide-range of bending rigidities~\cite{Saric2012}. This linear arrangement tends to align with the direction of largest curvature, eventually forming arcs and rings that completely surround the deformed vesicle~\cite{Koltover1999,Vahid2017}. Maybe even more surprisingly, spheres that interact with a pairwise repulsive interaction can form stable clusters on the membrane due to the non-additive forces induced by the membrane deformations~\cite{Kim1998}. Similarly, anisotropic membrane-deforming objects can assemble into different patterns, such as a linear and compact aggregates or disordered patterns~\cite{Dommersnes1999, Dommersnes2002, Olinger2016, Simunovic2013}. These intriguing observations from numerical calculations all originate from the non-additive nature of the multi-body interactions, but even the simplest case of three membrane-deforming particles has never been experimentally investigated.

We here employ an experimental model system consisting of Giant Unilamellar Vesicles (GUVs) and adhesive colloidal particles~\cite{VanDerWel2016,Sarfati2016} to study the configurations of three membrane deforming spherical objects. Our experimental setup allows us to single out membrane-deformation induced interactions. We quantitatively measure the particle positions in time in 3D using confocal microscopy and extract the free energy landscape for the three-particle case for the first time. To better understand the nature of this interaction, we carry out Monte Carlo simulations of three membrane-deforming colloids and sample the energy at various colloid distances. The simulations quantitatively agree with experiment, and allow us to identify the origin of the preferred particle arrangements to be the reduction in membrane bending energy in the neck region upon approach of the particles. The simplified particle shape and controlled membrane deformation by each particle allow us to draw conclusions about the non-additive nature of membrane-bending induced interactions.

\section*{Materials and Methods}

\subsection*{Chemicals} 
Phosphate-buffered saline (PBS) tablets,chloroform (99\%),styrene (99\%), itaconic acid (99\%),sodium phosphate (99\%), D-glucose (99\%), 4,4'-Azobis(4-cyanovaleric acid) (98\%, ACVA), N-hydroxysulfosuccinimide sodium salt (98\%,, Sulfo-NHS), 1,3,5,7-tetramethyl-8-phenyl-4,4-difluoro-bora-diaza-indacene (97\%, BODIPY)
deuterium oxide (70\%), and Bovine Serum Albumin (BSA) ($\geq$98\%)
 were purchased from Sigma-Aldrich; sodium azide (99\%) was obtained from Acros Organics; methoxypoly(ethylene) glycol amine (mPEG, Mw = 5000) from Alfa Aesar; 
1-Ethyl-3-(3-dimethylaminopropyl) carbodiimid hydrochloride (99\%, EDC) from Carl Roth;
NeutrAvidin (avidin) from Thermo Scientific;
1,2-dioleoyl-sn-glycero-3-phosphoethanolamine-N-[biotinyl(polyethylene glycol)-2000] (DOPE-PEG-biotin),
1,2-dioleoyl-sn-glycero-3-phosphocholine (DOPC),
1,2-dioleoyl-sn-glycero-3-phosphoethanolamine-N-(lissamine rhodamine B sulfonyl) (DOPE-Rhodamine),
from Avanti Polar Lipids and purchased from Sigma-Aldrich.
All chemichals were used as recieved. 
Deionized water obtained from a Millipore Filtration System (Milli-Q Gradient A10) with resistivity 18.2 M$\Omega~\cdot$ cm was used in all experiments.

\subsection*{Vesicle preparation}
Giant Unilamellar Vesicles (GUVs) were prepared using  a standard electroformation method. A lipid mixture containing 97.5wt.\% DOPC, 2.0 wt.\% DOPE-PEG2000-Biotin, and 0.5 wt.\% DOPE-Rhodamine was dissolved in chloroform to make a 1 g/L stock solution. Then, 10$\mu$l of this solution was deposited as a thin layer on two indium tin oxide (ITO) coated glass slides by spin coating. The thus coated glasses were dried for at least two hours in a desiccator at low vacuum to evaporate the organic solvent. After that, the slides were immersed in a solution of 100~mM glucose in 49:51 D$_2$O:H$_2$O. For electroswelling, an alternating electric field of 500~V/mm at 10~Hz was applied to the electrodes for 1.5~h, after which the frequency was reduced to 6~Hz to increase production efficiency for the next 30~min. Eppendorf tubes were BSA coated to reduce adhesion. To remove undesirable lipid structures such as tubes and small vesicles, the GUV solution was carefully mixed with 100 mOsm PBS and GUVs were let to sediment. After 10 min the supernatant was removed. 
Phosphate Buffer Saline was made by dissolving a PBS tablet in water to 100~mOsm. PBS would weight up by D$_2$O:H$_2$O to the equal weighted with particles and GUVs. Osmolarity was adjusted with an osmometer (GonoTec Osmometer Model 3000). All procedures were done at room temperature.

\subsection*{Particle preparation}
Polystyrene particles with a dense functionalization of carboxylic acid groups and 0.98 $\pm$ 0.03~$\mu$m diameter were coated with Neutravidin and mPEG 5000 following a protocol that was shown to yield a surface coverage of Neutravidin of $1.8\times10^4$~$\mu$m$^{-2}$~\cite{VanDerWel2017a}. Particles were dyed with fluorescent BODIPY dye throughout, by the following protocol of ref.~\cite{VanDerWel2017a}.
 
\subsection*{Microscopy and Optical Trapping}
An inverted Ti-E Nikon microscope equipped with a 60$\times$ water immersion (N.A.=1.2) objective lens  was used to capture confocal images with an A1R confocal head. Resonant mode was employed to capture (video) images of 512$\times$256 pixels at 15~kHz line scanning speed, equivalent to 59~frames/s. A 488~nm laser was used to excite the BODIPY dye inside the particles (depicted in green throughout) and the emission was detected between 500 and 550~nm. Simultaneously, a 561~nm laser was used to excite the rhodamine dye attached to a small fraction of lipids (0.5 wt.\%, depicted in magenta) and the emission was collected between 580 and 630nm. The microscope stage is mounted on a Madcity LAB nano Piezo z-positioner that can move the stage in the range of a hundred micrometers with high precision and speed. Images were acquired in two separate and independent channels. 

We used home-built optical tweezers to bring the particles to the top of the vesicle at the start of each experiment. Our optical tweezers consisted of a highly focused laser beam (Laser QUANTUM Opus with $\lambda=$1064~nm) that was integrated into the confocal microscope through the fluorescent port and merged into the microscope light path with an IR short pass filter. The same objective was used for confocal imaging and focusing the trapping beam inside the sample; the correction collar of the objective was set slightly lower than the real thickness of the coverslip to reduce the spherical aberration~\cite{Reihani_2011}.

\subsection*{Sample and chamber preparation}
Glass coverslips were coated with BSA then washed three times with PBS buffer. We also tested a polyacrylamide coated coverslip and did not observe a significant difference in the result. Throughout this work, BSA-coated coverslips were used. GUVs were mixed gently with colloidal particles dispersed in isotonic PBS buffer to prevent spontaneous tube formation on the membrane. Then the mixture was injected into a microscope sample holder. The vesicles were let to sediment to the cover glass before positioning the colloidal particles and imaging. We achieved a decrease in the GUV tension which is needed for particle wrapping by keeping the sample holder open. The evaporation of water leads to a gradual increase of the ion concentration of the outer solution, which creates an osmotic pressure difference between the inside and outside of the GUV and hence a decrease in the GUVs' tension.

We excluded non-spherical GUVs from our measurements for two reasons: (1) there might be an anisotropic force due to the elongation of membrane~\cite{Vahid2017} and (2) the higher symmetry simplifies the analysis in that it reduces the problem from $\theta=[0,2\pi]$ to $\theta=[0,\pi/2]$ and hence requires less data.
 
\subsection*{Image analysis and Particles tracking}
Particle position in the acquired (video) images were obtained using the Python-based software package Trackpy 0.5.0~\cite{Allan2021Soft-matter/trackpy:V0.5.0}. The software traces fluorescent particles by finding their center of mass using a Gaussian fit in X and Y with sub-pixel resolution. At the start of each experiment, the membrane size, position and tension was extracted from a 3D stack of images using a python routine that finds the vesicle center, contour, and fluctuation spectrum\cite{Pecreaux2004} with subpixel precision~\cite{vanderWel2016CircletrackingV1.0}.
The particles' confinement to the membrane enabled us to retrieve their 3D position from their two-dimensional X-Y coordinates~\cite{VanDerWel2016,Azadbakht2023a}. 

\subsection*{Simulations}

Coarse grained Monte Carlo simulations of the membrane-colloid system were carried out using a previously developed triangulated fluid membrane model \cite{Saric2012,Saric2012a,Saric2013,VanDerWel2016}. This consists of a network of triangles with 5882 vertices that dynamically undergo Monte Carlo moves involving translation of vertices and edge-swapping. Each move is associated with an energy change calculated from the angles between the normals of the triangles and their area, which together with the edge-swapping can replicate the bending rigidity, tension and fluidity of biological membranes. Each move is accepted according to the Metropolis-Hastings algorithm, satisfying detailed balance. Between the vertices there is a hard-core volume exclusion where the minimum distance between vertices is $D_{\mathrm{mem}}$. We choose a membrane bending rigidity $\kappa = 15 k_{\mathrm{B}}T$ and membrane tension $\sigma \approx 1 k_{\mathrm{B}}T/D_{\mathrm{mem}}^{2}$. The colloids are modelled using a hard core volume exclusion with the membrane and each other combined with an attraction to the membrane. The attraction takes the form $V_{\mathrm{att}} = -\epsilon\left(\frac{D_{\mathrm{mem}} + D}{2d}\right)^{6}$ where $d$ is the distance between the centre of the colloid and a membrane vertex, $D = 5D_{\mathrm{mem}}$ is the diameter of the colloid and $\epsilon$, the strength of the interaction, is chosen to be 7k$_{\mathrm{B}}$T to ensure full and tight wrapping of the colloids by the membrane. The attraction is cut off at $d_{\mathrm{cut}} = 1.5 \frac{D_{\mathrm{mem}} + D}{2}$, after which it is zero. In contrast to earlier work where the colloids are not fully wrapped by the membrane, \cite{Saric2012, Bahrami2018, Reynwar2011a} the colloids here are close to 100\% wrapped with a very tight neck (see Fig. \ref{fig:Fig.5}b top panel) which closely resembles the experimental set-up. Indeed we confirm the theoretically predicted behaviour of a discontinuous transition from semi wrapped to fully wrapped as adhesion energy per colloid area is increased \cite{Raatz2014}. To measure the potential energy for different configurations of the three colloids, we tether each colloid using a stiff harmonic spring at a chosen fixed angle with respect to the centre of the vesicle, and vary these angles to sample different configurations. The colloids are free to move in the radial direction and can be wrapped by the membrane. We run 20 seeds (with different randomised Monte Carlo realisations) per fixed distance between the 3 colloids and obtain an averaged potential energy profile. We eliminate variable binding energy by first creating bins of data with the same total binding energy, averaging each bin independently and then plotting the mean of these averages. When we plot the bending energy of the membrane as a function of arc distance, we average the membrane bending per bead for a strip which is $10 D_{\mathrm{mem}}$ wide and follows the arc that is defined by either $r$ or $\ell$ (see Figure \ref{fig:Fig.5}). Simulations are performed in the canonical ensemble. 

\section*{Result and Discussion}

\begin{figure}[hbt!]
\centering
\includegraphics[width=1\linewidth]{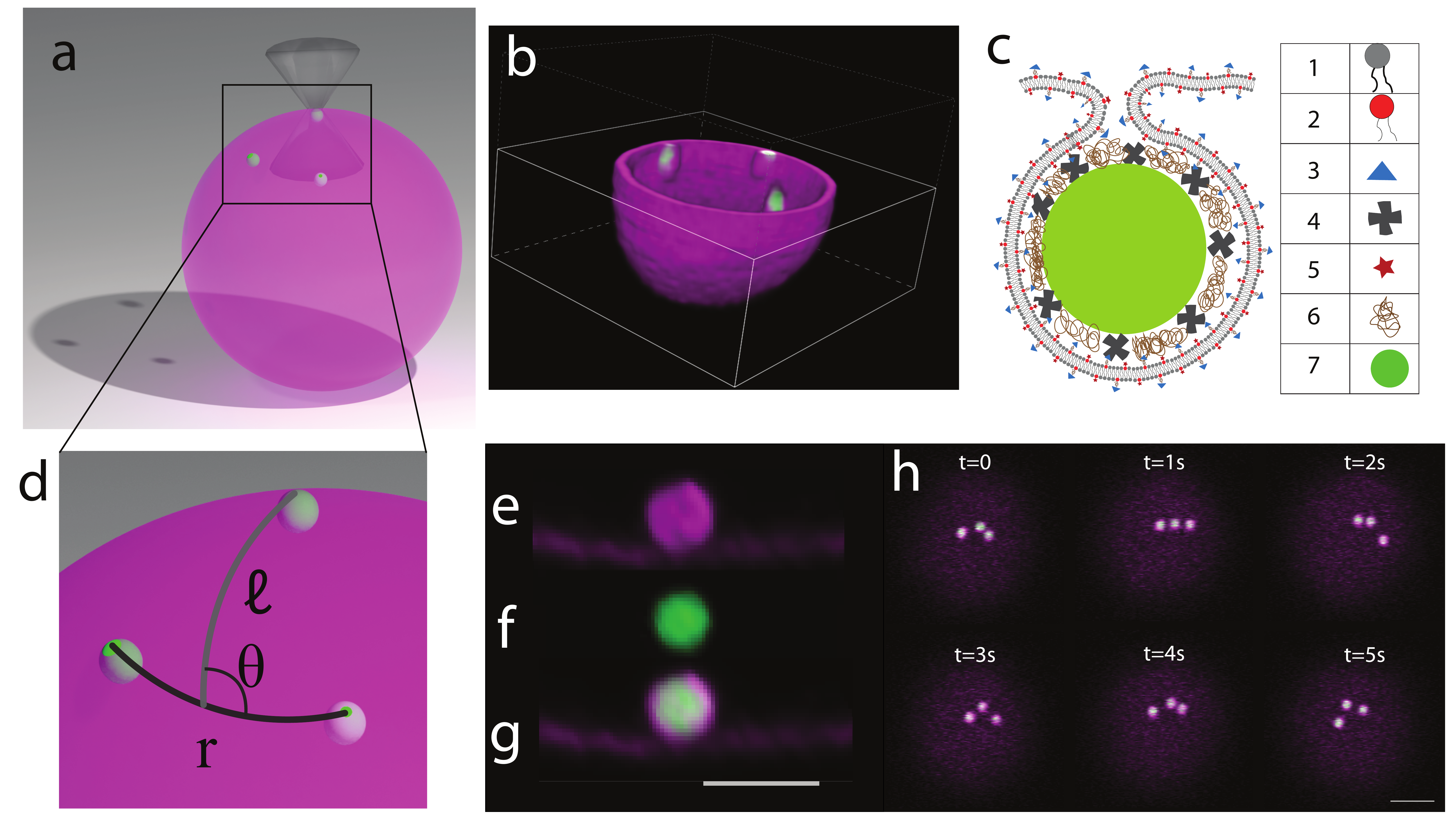}
\caption{\textbf{Experimental setup for measuring membrane-mediated interactions between three spherical, membrane deforming particles.} (a) Schematic of the model system that consists of Giant Unilamellar Vesicles (GUVs, depicted in magenta) and  membrane-wrapped polystyrene particles (depicted in green), three of which are positioned on top of the vesicle using optical tweezers. 
(b)Three-dimensional reconstruction of a confocal image of a hemisphere of a $20\mu m$ diameter GUV with three wrapped particles.
(c) Detailed schematic of particle and membrane functionalization, and membrane-wrapped configuration (not to scale). 1-DOPC lipid 2-DOPE lipid 3-Biotin 4-NeutrAvidin 5-Rhodamine 6-polyethylene glycol (PEG) 7-Polystyrene particle. (d) Depiction of the definition of the parameters used to describe particle arrangement on the membrane.
(e-g) Membrane wrapping of colloids can be identified using different dyes and separate fluorescent detection channels, e) membrane fluorescence is detected between 580-630nm, and f) particle fluorescence between 500-550nm;  g) an overlay of both channels allows identification of overlap of membrane and particle fluorescence by the white color. Scale bar is 2$\mu$m.  (h) Stills from an image sequence of three wrapped particles on a GUV. Scale bar is 5$\mu$m. Time $t$ is indicated in the stills.}
\label{fig:Fig.1}
\end{figure}

To quantitatively investigate how three membrane deforming objects interact, we use an experimental model system consisting of model lipid membranes realized by GUVs with diameters ranging between 20-25~$\mu m$ and membrane-adhering and deforming colloidal particles~\cite{VanDerWel2016}. We induce adhesion between the colloids and the GUV by employing $0.98~\pm 0.03\mu m$ diameter polystyrene spheres functionalized with $1.8\times 10^4$ $\mu$m$^{-2}$ Neutravidin and equipping the GUVs with 2\%w/w biotinylated lipids, see Methods for details and Fig.\ref{fig:Fig.1}a,b,c. Once attached to the membrane, the colloids can become wrapped by the membrane. This happens for sufficiently low surface tensions $\sigma$ and if the adhesion energy per colloid surface area $u_{ad}$ is larger than the bending energy $E_b$ required to deform the membrane with bending rigidity $\kappa$~\cite{VanDerWel2016,Spanke2020,Spanke2022,Azadbakht2023a}. We achieve the latter by leaving our sample holder open such that gradual evaporation of water, which slowly increases the osmolarity of the outer fluid leads to a deflation of the vesicles. This deflation causes the membrane tension to lower to a level below 10~$n$N/m. Wrapped particles induce a spherically symmetric membrane deformation and can interact through this deformation~\cite{VanDerWel2016,VanderWel2017}.

We validate that colloids have been wrapped by the membrane with confocal microscopy. To this end, lipid membranes are labelled using rhodamine-B (shown in magenta, see Fig.\ref{fig:Fig.1}e), and colloids are dyed with BODIPY (shown in green, see Fig.\ref{fig:Fig.1}f), we imaged both GUVs and colloids simultaneously in separate fluorescence channels. Overlapping signal of membrane and colloids leads to a white color (Fig.\ref{fig:Fig.1}g). The position of the colloidal particles with respect to the vesicle membrane allows clear distinction between wrapped and unwrapped particles, see Fig.\ref{fig:Fig.1}b and g. Instead of relying on diffusion-based random adhesion to the lipid membrane only, we also employ optical tweezers to bring the colloidal particles in contact with the GUV. This allows us to not only speed up the adhesion process of particles on the GUV, but also their wrapping by the membrane~\cite{Azadbakht2023a}.

To simultaneously and at high speed image the conformations and interactions of three membrane-deforming colloidal particles by confocal microscopy, the particles need to be at roughly the same focal depth (Fig.\ref{fig:Fig.1}h and Movie S1). We achieve this by positioning three wrapped particles on top of the vesicle (Fig.\ref{fig:Fig.1}a,h) at the start of each experiment. After releasing the confinement imposed by the optical tweezers, we extracted their x- and y-position in time using Python based image analysis routines. Knowledge about the GUV size and position and the wrapping state allows us to infer the z-position of the colloidal particle from their x- and y-coordinates, and hence use the much faster 2D imaging.

We characterize the relative positions of the three interacting particles using three parameters: (1) $r$ is defined as a side of a triangle made from center of three particles, (2) $\ell$ is defined 
geodesic line from the middle of $r$ to the third particle,  and (3) $\theta$ is defined as the angle between $r$ and $\ell$, schematically depicted in Fig.\ref{fig:Fig.1}d.

\begin{figure}[hbt!]
\centering
\includegraphics[width=1\linewidth]{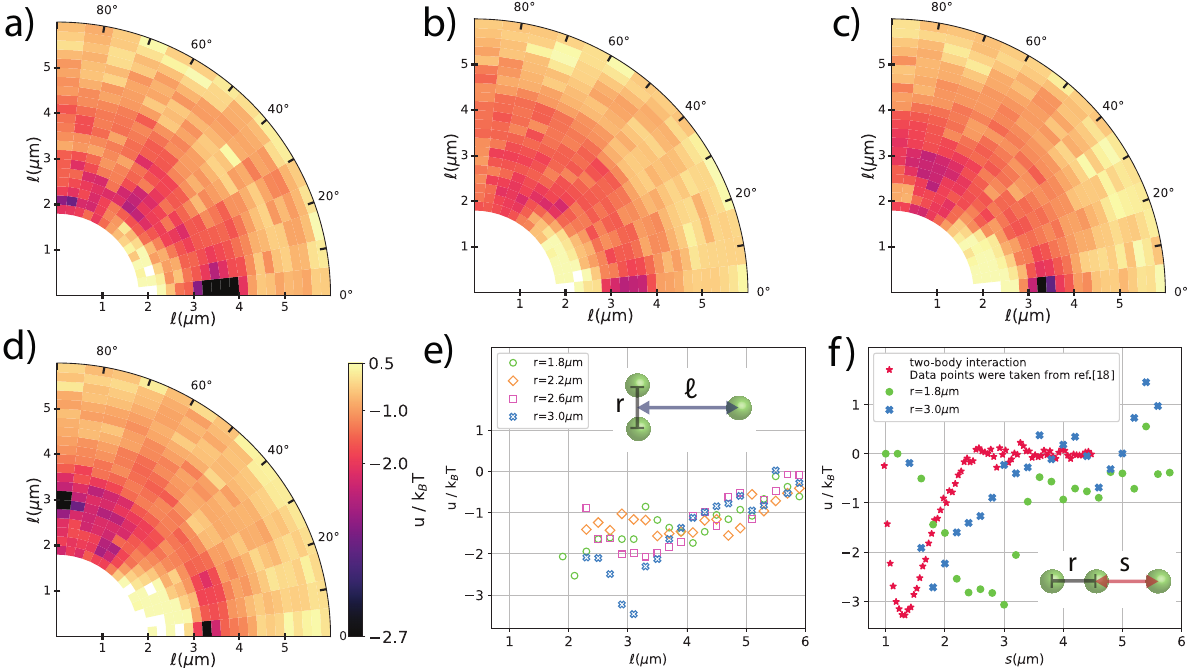}
\caption{\textbf{Free energy $u$ for three membrane-deforming spheres} as a function of the distance $\ell$ and angle $\theta$ as defined in Figure \ref{fig:Fig.1} where two particles are located at $\ell=\pm r/2$ on the horizontal axis. Polar plot of the free energy for
(a) $r=1.8 \pm 0.2 \mu m$;
(b) $r=2.2 \pm 0.2 \mu $m; 
(c) $r=2.6 \pm 0.2 \mu $m; and
(d) $r=3.0 \pm 0.2 \mu $m; 
(e) Free energy in k$_{\mathrm{B}}$T as a function of $\ell$ for the triangular arrangement, i.e. $84^\circ <\theta < 90^\circ$; 
a switch from a linear arrangement of the three spheres to an equilateral triangle at larger values of $r$ is visible. 
(f) free energy in k$_{\mathrm{B}}$T as a function of distance $s$ defined as shown in the inset schematic, for the three particles to be arranged in line, i.e. $0^\circ<\theta<6^\circ$ (blue squares) and for $r=1.8\pm 0.2 \mu$m (green circles) and $r=3.0\pm 0.2 \mu$m; in addition, the free energy for two membrane deforming spheres is shown (red stars, taken from~\cite{VanDerWel2016}) for comparison.
}
\label{fig:Fig.2}
\end{figure}

To extract the free energy associated with different particle arrangements on the membrane from their positional data different methods are available in the literature. Some of these methods require a long continuous trajectory ~\cite{Frishman2020,Gnesotto2020}, which makes it not possible to use them for such experiments where the particles might move out of the microscope's focus and disappear from the field of view. Maximum-likelihood analyses of particle trajectories circumvent this problem, but require an assumption of the three-body interaction potential and diffusion coefficient in advance~\cite{Sarfati2017}. Here, we therefore used a standard Boltzmann statistics approach that only requires us to obtain information about the particles' position albeit at the cost of large amounts of data. We split the data according to a distance between any two particles, $r$, into bins of $\pm 0.2~\mu$m in a range from $r~=~1.8...3.0~\mu$m. This bin size ensures good statistics by containing each at least 22.000 data points while at the same time being reasonably small with respect to the microscope resolution. This approach is equivalent to considering a probe particle that explores the interaction with two particles located within a fixed distance $r\pm \delta r$. Moreover, $\theta$ can vary from 0$^\circ$ to 360$^\circ$, but all data can be mirrored to one quarter due to its symmetry (Fig.\ref{fig:Fig.2}a).

Using this approach, we infer the free energy of three particles for four values of the distance between any two particles, $r$, and plot them as a function of the distance $\ell$ and angle $\theta$ of the third particle with respect to the center the two other particles, see schematic shown in Figure~\ref{fig:Fig.1}d. 
The color scale was scaled as a power-law with power 0.3 for better visual presentation, with darker colors corresponding to a lower free energy. For the smallest value of $r=1.8 \pm 0.2 \mu$m, the most likely arrangement is linear with one particle being located at a distance of about 3.5 $\pm$ 0.6~$\mu$m from the center of the two other particles (Fig. \ref{fig:Fig.2}a). Intriguingly, the distances between the three particles in the linear configuration are larger than those previously found for two particles, which was about 1.25~$\mu$m~\cite{VanDerWel2016}. Additionally, the maximum strength of the interaction between two particles was previously measured to be -3.3~k$_{\mathrm{B}}$T, which is similar to the value of -3.0~$\pm$~0.2~k$_{\mathrm{B}}$T that we find for the linear three-body configuration here. Arrangements with smaller values of $r$ were so infrequently observe that we were unable to measure the free energy with high precision and hence we do not report them here. Thus, our data suggest that the linear arrangement occurs at slightly larger distances and nearly equal interaction strength compared to the two-body scenario. As $r$ increases to 2.0 $\pm$ 0.2 $\mu$m, the linear configuration of the three spheres is still the most likely state, Fig. S1 a, and persists even for larger values of $r$=2.2 $\pm$ 0.2 $\mu$m (Fig.\ref{fig:Fig.2}b) and $r$=2.4 $\pm$ 0.2 $\mu$m (Fig.S1~b).

A second preferred configuration appears when moving the two particles even further apart, to $r=2.6 \pm 0.2 \mu$m and $r=2.8 \pm 0.2 \mu$m, see Figure \ref{fig:Fig.2}c and Figure S1~c. This arrangement is equivalent to a triangular arrangement of the three spheres.  
Initially the second minimum corresponding to the triangular configuration is shallow with respect to k$_{\mathrm{B}}$T, allowing particles to easily escape. Upon a further increase of the distance between the two particles, however, the minimum belonging to the linear arrangement ($\theta=0$) becomes less dominant and the triangular arrangement at $r$ and $\ell \simeq 3.0\mu $m and $\theta \simeq 90^\circ$  becomes energetically preferred, albeit by a small difference (Fig.\ref{fig:Fig.2}d and f). This becomes more clear, when we plot the free energy for the triangular state as a function of the distance $\ell$ of the third particle for all distances $r$, see Figure~\ref{fig:Fig.2}e. Only for the largest distance $r=3.0\pm 0.2 \mu$m, the equilateral triangle is preferred, with a minimum free energy of about -3.2$_{\mathrm{B}}$T at about $\ell=3.0\pm 0.4 \mu$m. However, the minimum has has a broader width than the linear configuration. While a free energy minimum was found for all values of $r$ in the triangular state, the most pronounced minimum is that for $r=3.0\pm~0.2~\mu$m, see Figure~\ref{fig:Fig.2}e. In this state, the three spheres are arranged in an equilateral triangle: two spheres have a distance $r=3.0\pm 0.2 \mu $m, which for a perfect equilateral triangle would require $\ell=2.6\pm 0.2\mu$m. We find $\ell \simeq 3.0\pm 0.4\mu$m from the experiments, which agrees within the error with an equilateral triangle configuration.

Our experimental observations of a linear arrangement and equilateral triangle of three membrane deforming spheres on a closed lipid membrane have never been reported before in experiments. They are in line, however, with the observation of hexagonal arrangements found by Ramos et al. for negatively charged particles physisorbed onto oppositely charged surfactant vesicles~\cite{Ramos1999}. Besides that the vesicle is not made from lipids, in the same work, irreversible aggregation of colloidal particles was reported - in contrast to our observations here and earlier~\cite{VanDerWel2016}, and it is unclear if these particles interacted purely through deformations. For our system, irreversible aggregation only occurred in the presence of spurious lipid structures or for combinations of wrapped and non-wrapped particles~\cite{VanderWel2017}.

Our observations of a coexistence of linear an triangular arrangements on closed spherical lipid membranes have also never before been reported in analytical or numerical work. In fact, there are only a few reports in the literature that consider many objects that interaction through deformations on spherical membranes~\cite{Bahrami2012,Saric2012}, because most assume flat sheets for simplicity~\cite{Park1996,Kim1998}. Yet, this seemingly straightforward simplification may lead to different results, as was shown for two-body interactions~\cite{Vahid2016, Idema2019a}. 
Being aware of this, and keeping in mind that these works predicted a repulsion between two deformations, we still would like to point out that early analytical work on three membrane-deforming objects on tensionless flat sheets found that their interaction depended on their precise arrangement and that the free energy was predicted to have maximum repulsion for an equilateral triangle~\cite{Park1996,Yolcu2012}.

Previous numerical work that most closely captures the physics of our experimental model system considered adhesive nanoparticles on flat and spherical membranes~\cite{Saric2012}. These numerical calculations found a preference for linear arrangements at bending rigidities ranging from 10-100~k$_{\mathrm{B}}$T, which is the regime our GUV membrane falls into. For lower and higher bending rigidities, hexagonal arrangements were found. This preference was attributed to the higher gain in adhesion energy compared to the bending energy costs for the linear arrangement compared to the hexagonal one. Free energy calculations for three particles on flat membranes with two particles in close contact and the third one coming in linearly or in a triangular orientation confirmed a preference for the linear arrangement. However, larger values for $r$ were not considered, which  might have shown both the linear and the triangular state we observe in the experiments. In addition, the nanoparticles considered in this work were never found to be fully wrapped by the membrane but rather to indent the membrane at maximum to about half their size.

To better understand the effect of an additional object on the two-body interaction as well as the non pairwise nature of the interaction, in Figure~\ref{fig:Fig.2}f we plotted the free energy of the three particles arranged in a line in a different way, namely as a function of $s=\ell-\frac{r}{2}$. This allows us to compare our data more easily to the two-body interaction data measured previously in ref.~\cite{VanDerWel2016}. This way of presenting the data is equivalent to considering the two-body interaction in the presence of a third particle, when all particles are arranged on a line, since the direction of the membrane-deformation induced forces is parallel to this line. Interestingly, at short distances of $r~=~1.8\pm~0.2 \mu$m, the interaction free energy with the third particle is clearly not simply an addition of the interactions with two spheres at a distance $s$ and $s+r$. Instead, the magnitude of the interactions is comparable to what has been found for two  spherical particles that deform the membrane in a similar way. Secondly, in contrast to pure two-body interactions, the minimum of the interaction is shifted to larger distances $s$ and features a broader interaction range with roughly 1.2~$\mu$m, see Fig.~\ref{fig:Fig.2}f. When the third particle is further away, i.e. for r=3.0~$\mu$m, the minimum of the interaction potential moves to smaller values and closer to the value found for pure two-body interaction. In the limited of an infinitely distant third particle, the two-body interaction potential is expected to be recovered. Therefore, we find that the effect of a third particle on the interaction of two particles is to perturb the equilibrium distance between them. This is stronger the closer the third particle is to the interacting pair. This finding clearly highlights and confirms experimentally for the first time the predicted non-additive nature of the interaction.

To investigate these interactions further, we performed Monte Carlo simulations consisting of a triangulated fluid membrane and adhesive colloids (see Methods for details). We focus on interactions where each colloid wraps individually and then interacts with other individually wrapped colloids. Due to the thermal noise in the system, it is challenging to measure interactions of the order of a few k$_{\mathrm{B}}$T found in experiment. The biggest source of noise measured in the energy in simulations is from variable colloid-membrane binding. When just one membrane bead binds or unbinds from a colloid in the wrapped state, there is a large difference in the energy. To mitigate this, we choose to measure the potential energy at constant binding. This is likely reflective of the experiments, where binding is irreversible due to the strong interaction between the biotin-strepatividin linkers. We compare the potential energy differences measured in simulation to the free energy differences measured in experiment. The simulations reproduce the energy minima found in experiment remarkably well, suggesting that the interactions measured in experiment are neither entropic nor due to variable binding.

\begin{figure}[hbt!]
\centering
\includegraphics[width=1.\linewidth]{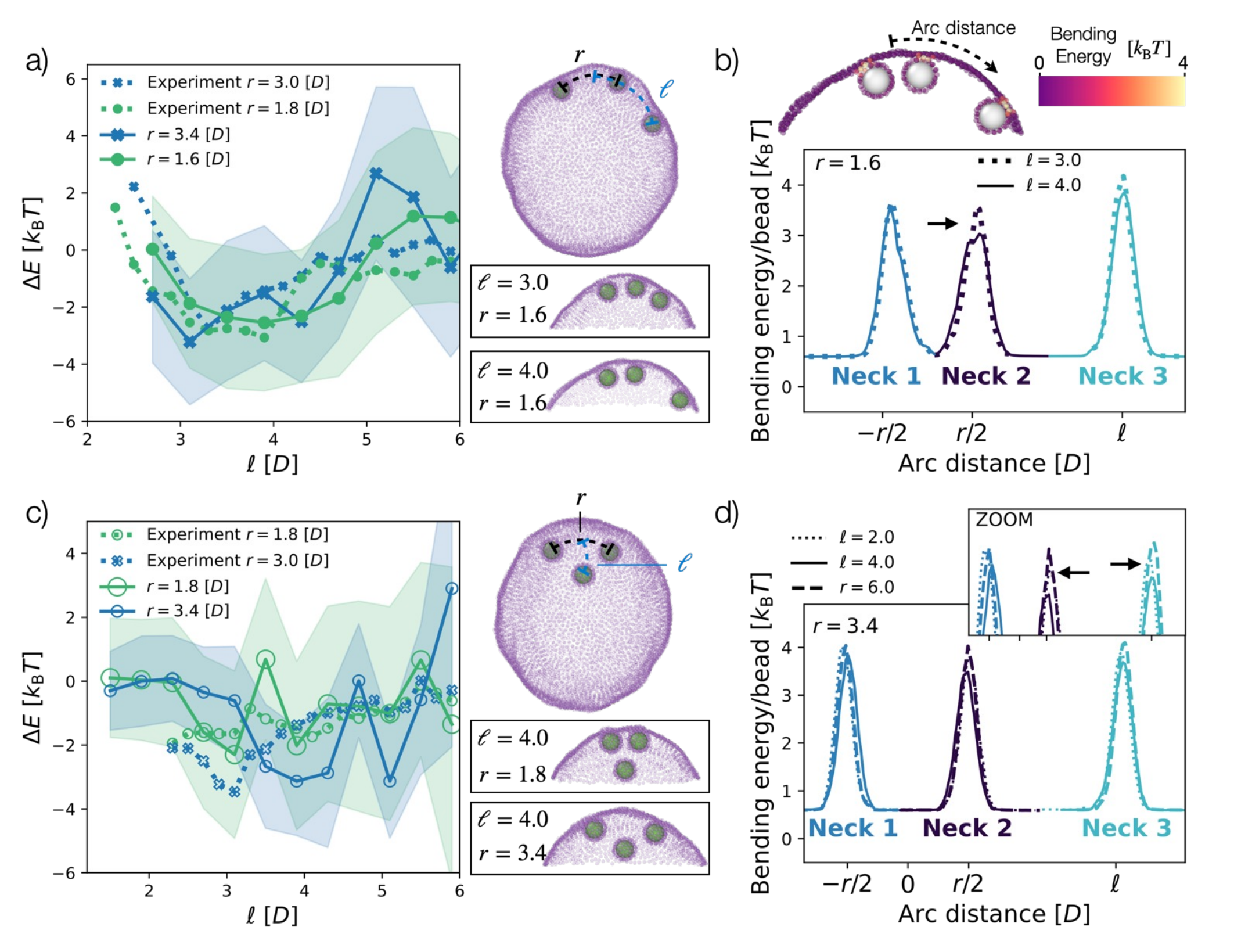}
\caption{\textbf{Simulations for three colloid interactions.} Measurements of the change in bending and stretching energy ($\Delta E$) as a function of $\ell$ for a) and b) for the linear arrangement and c) and d) for the triangular arrangement. The snapshots in a) and c) show how $r$ and $\ell$ are measured in both the linear and triangular configuration. a) shows that in the linear regime there is a minimum at fixed $\ell \approx 3.0 - 4.0~D$, compared to large ($\ell = 6.0~D$) or small ($\ell = 2.5~D$) values of $\ell$, for a range of $r$ in both simulation and experiment, where $D$ is the diameter of the colloid. c) shows there is a single minimum at fixed $\ell \approx 4.0~D$ for $r = 3.4$, compared to large ($\ell = 6~D$) or small ($\ell = 2.0~D$) values of $\ell$, in simulation and a single minimum at fixed $\ell = 3.0~D$ and $r = 3.0$ in experiment for the triangular regime. Bending energy per membrane bead as a function of distance to the midpoint between the first two colloids are shown in b) and d) for the linear and triangular state, respectively. We average the energy per membrane bead of a strip of membrane beads, $10~D$ wide, along the arcs defined by $r$ and $\ell$ (not including the colloid-bound membrane beads). This allows us to visualise the three membrane necks as three peaks in energy-distance space. In b) it is clear that the energy of the middle membrane neck is lower for the minimum of $\ell = 4.0$, $r=1.6$. In d) the neck energy reduction is less pronounced, but the zoomed inset shows that the energy minimum comes from a reduction in the neck energies for intermediate $\ell = 4.0D$.} 
\label{fig:Fig.5}
\end{figure}

Figure \ref{fig:Fig.5}a shows simulation results alongside experimental results for the interaction between three colloids in the linear arrangement. The difference in bending energy and energy due to surface tension, $\Delta E$, is plotted as a function of the distance between the third colloid and the midpoint between the first two colloids, $\ell$ (see Figure \ref{fig:Fig.5}a top right snapshot). We find a minimum for $r = 3.4~D$ and $r = 1.6~D$ at fixed $\ell \approx 3.0 - 4.0 ~ D$ which reflects the findings from experiment for $r = 3.0~D$ and $r = 1.8~D$, respectively.  This favourable configuration is shown in the bottom snapshot in Figure \ref{fig:Fig.5}a. Figure \ref{fig:Fig.5}c shows the simulation and experiment results for the triangular state, where $\Delta E$ is plotted as a function of the same parameter, $\ell$ (see Figure \ref{fig:Fig.5}c top right snapshot). This interaction was more difficult to capture in simulations, as can be seen by the noise in the interaction profile. However, we see a pronounced minimum, for $r = 3.4 D$ and $\ell = 4.0 D$ which probably reflects the only minimum found in experiment at slightly smaller $r = 3.0 D$ and $\ell = 3.0 D$.

The advantage of these simulations is that we can look into where the interactions are coming from. To begin to explore this, we first revisited a system with only two membrane-deforming colloids. As can be seen in Figure S2, the potential energy is minimal for the closest colloid-colloid distance we could simulate, $s=1.5D$, and dominated by differences in bending energy, reflecting experiments and confirming previous simulation results \cite{VanDerWel2016}. To interrogate the source of this interaction, we plot the local bending energy of the membrane as a function of arc distance from the center of the two colloids for different colloid configurations (see Methods for details). From this we find that the tightest, highly bent part of the neck has a lower bending energy at close colloid-colloid distances.

Turning to the case of three colloids we find analogous results. We again plot the average bending energy per membrane bead as a function of arc distance, see Figures \ref{fig:Fig.5}b and d. In the linear regime, the two closest colloids have a lower membrane bending energy at the neck than the membrane neck of the furthest colloid, see Neck 1 and Neck 2 vs Neck 3 in Figure \ref{fig:Fig.5}b, directly reflecting the two colloid interaction. There is another reduction in the membrane neck bending energy of the middle colloid when the third colloid is an intermediate distance away, see Neck 2: $\ell = 4.0 D$ vs $\ell = 3.0 D$ Figure \ref{fig:Fig.5}b. The bending energy differences between configurations in the triangular state also comes from a reduction in bending energy of the tightest part of the neck, this time for all three necks, see Figure~\ref{fig:Fig.5}d. We expected to find these interactions mediated by the membrane area between the necks, but instead we find that it is mediated by the tightest part of the membrane necks. The two colloid and three colloid interactions have the same source, a reduction in bending of the membrane necks. Yet the three colloid potential minimum is for colloid-colloid distances which are larger than the distance of the two colloid potential minimum, highlighting the non-additive nature of this interaction.

\section*{Conclusion}
We employed a model system consisting of colloidal particles and GUV's to measure the membrane-mediated interactions between three membrane-deforming, spherical objects. We found two distinct configurations for three-particle on the membrane driven by minimization of the membrane bending energy: a linear arrangement, and an equilateral triangular configuration at slightly longer distances. The minimum of the free energy was in both cases found to be around -3 k$_{\mathrm{B}}$T. We confirmed and quantitatively investigated the non-additive interaction of the particles on spherical membranes for the first time, and found that the third particle does not enhance the interaction but instead pushes the minimum towards a larger distance.

Extrapolating beyond two and three membrane deforming objects, the simplicity of our observations of only two states, linear and equilateral triangular, and their high symmetry in the particle arrangement suggests that many-body arrangements might either be hexagonal lattices or lines, or a combination of both. This is similar to early, albeit different,  because possibly not based on membrane-bending mediated interactions only, experimental work \cite{Ramos1999} and numerical predictions on flat \cite{Park1996} and spherical membranes \cite{Saric2012}. We stress, however, that this is a hypothesis only that needs to be tested, either in experimental model systems or numerical simulations.

\section*{Author Contributions}
AA and DJK designed the experiment. AA carried out the experiments and analysis of the data. DJK conceived the experiments and supervised experiments and data analysis. BM and A\v{S} designed the simulations. BM and JM carried out the simulations and A\v{S} supervised the simulations. AA, BM, A\v{S} and DJK contributed to writing the article. 

\section*{Acknowledgments}
We gratefully acknowledge useful discussions with Casper van der Wel, help by Yogesh Shelke with PAA coverslip preparation, and support by Rachel Doherty with particle functionalization. AA and DJK would like to thank Timon Idema and George Dadunashvili for initial attempts to simulate the experimental system. DJK would like to thank the physics department at Leiden University for funding the PhD position of AA. B.M. and A.Š. acknowledge funding by the European Union’s Horizon 2020 research and innovation programme (ERC Starting Grant No.~802960).

\section*{Supplementary Material}

An online supplement to this article can be found by visiting BJ Online at \url{http://www.biophysj.org}.

Movie S1: Movie showing an experimental measurement of three wrapped spheres on the top of a GUV, recorded with a confocal microscope at 59 frames per second and 4$\times$ accelerated.

\bibliography{references}

\end{document}


\date{}
\renewcommand{\figurename}{Movie}

\newpage
Movie is available online. Here, an image of the
video is shown.
\begin{figure}
  \centering
  \includegraphics[width=1\linewidth]{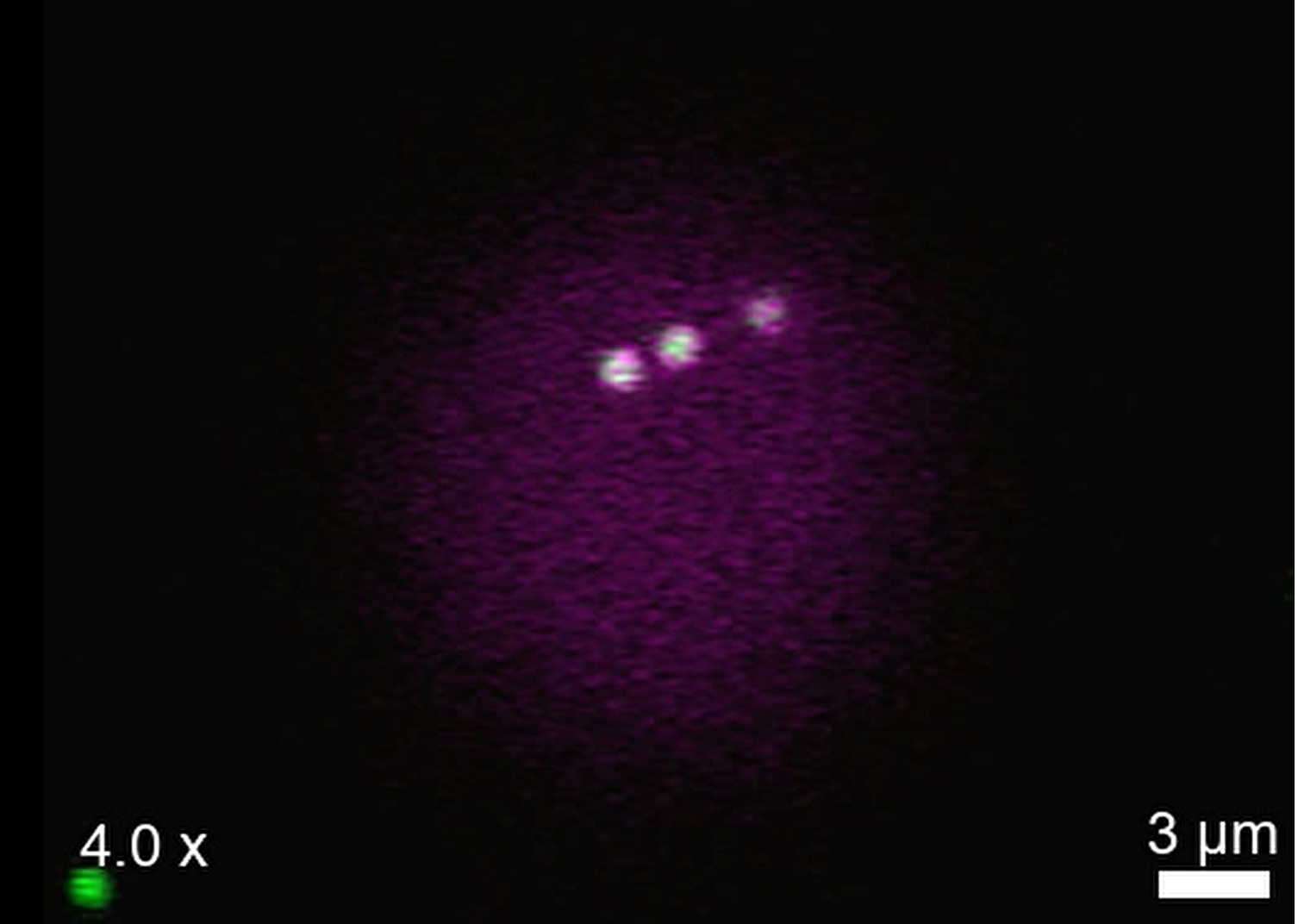}
  \caption{Movie showing an experimental measurement of three wrapped spheres on the top of a GUV, recorded with a confocal microscope at 59 frames per second and 4$\times$ accelerated.}
\end{figure}

\renewcommand{\figurename}{Figure}
\setcounter{figure}{0}

\newpage

\begin{figure}[hbt!]
\centering
\includegraphics[width=1\linewidth]{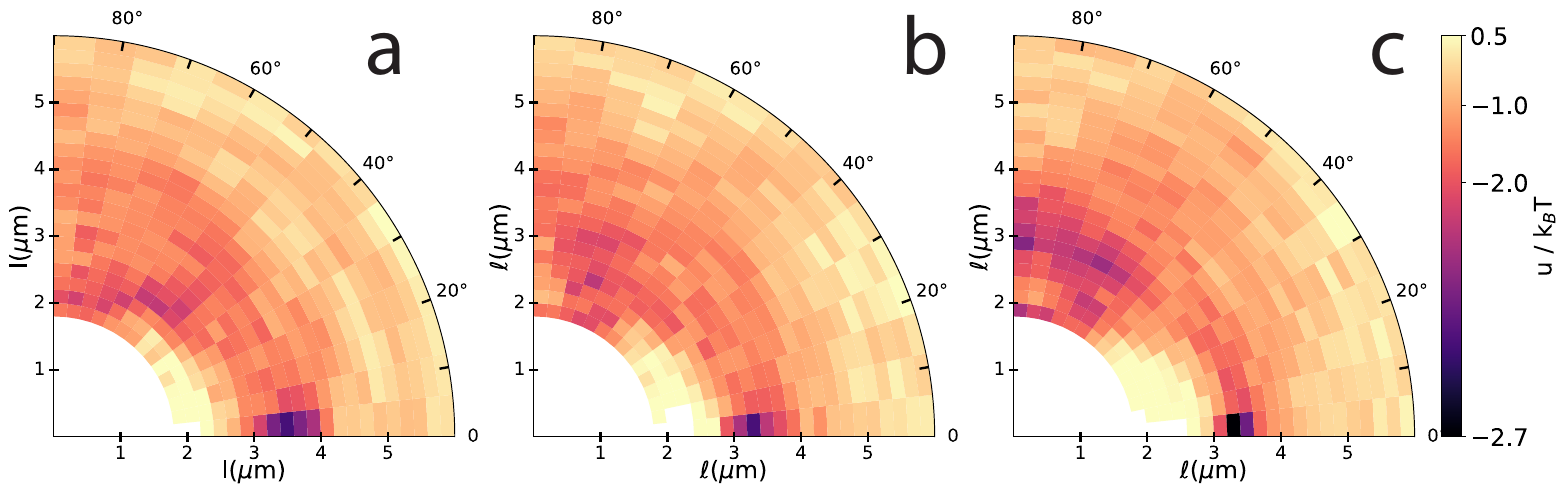}
\caption{\textbf{Polar plot of the free energy} of three membrane-deforming spheres for distance $\ell$ and angle $\theta$ as defined in Figure 1d where two particles are located at $\ell=\pm r/2$ with respect to the origin and a) $r=2.0 \pm 0.2 \mu m$, b) $r=2.4 \pm 0.2 \mu $m,
c)$r=2.8 \pm 0.2 \mu m$ 
A switch from a linear arrangement of the three spheres to an equilateral triangle at larger values of $r$ is visible. 
}
\label{fig:Fig.SI1}
\end{figure}

\newpage

\begin{figure}[hbt!]
\centering
\includegraphics[width=1.\linewidth]{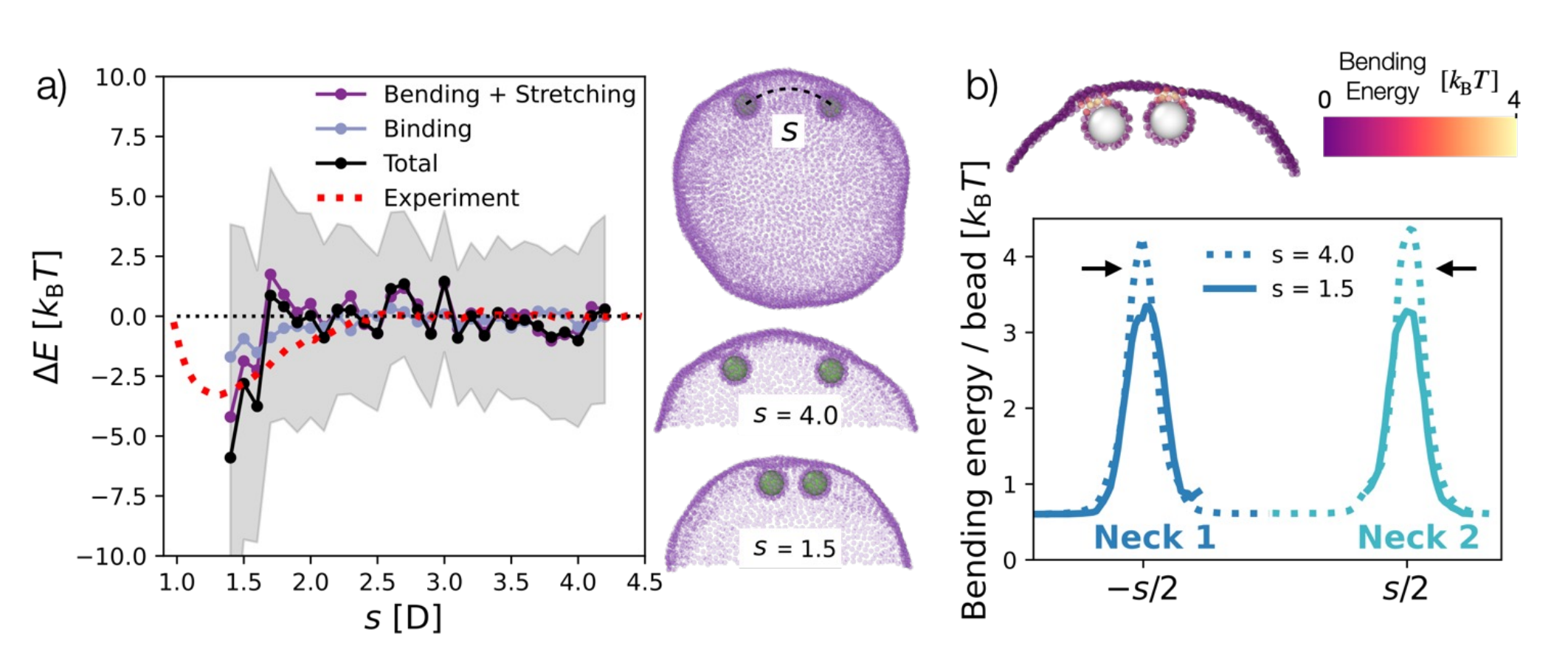}
\caption{\textbf{Simulations for the two colloid interaction} The interaction between two membrane-wrapped colloids measured in simulations (black dotted line) and compared to experiment (red dotted line, taken from van der Wel \cite{VanDerWel2016}) is shown in a). The interaction in simulation is dominated by the difference in bending energy, which is low for small $s$ compared with large $s$. We plot the membrane bending energy per membrane bead in b) for a strip of membrane following the arc defined by $s$, and excluding colloid-bound membrane beads. This can be visualised by colour (see b), top panel). The peaks in membrane bending correspond to the necks joining the colloid-bound membrane to the membrane making up the vesicle bulk. It is clear that the membrane neck has a lower bending energy per bead when the colloids are close, i.e. for $s/D=1.5$ instead of $s/D=4.0$ (see b), bottom plot, black arrows).} 
\label{fig:Fig.S1}
\end{figure}

\bibliographystyle{unsrt}
\bibliography{references}